\documentclass[12pt,a4paper]{iopart}
\usepackage{graphics,amssymb}
\usepackage{graphicx}
\usepackage{subfigure}
\usepackage{color}
\begin{document}
\title{Quantum noise properties of multiphoton transitions in driven
  nonlinear resonators}
\author{V. Leyton$^1$, V. Peano$^2$, and M. Thorwart$^1$}
\address{$^1$I.\ Institut f\"ur Theoretische Physik, Universit\"at
  Hamburg, Jungiusstra{\ss}e 9, 20355 Hamburg, Germany
  \\
  $^2$Department of Physics and Astronomy, Michigan State University,
  East Lansing, MI 48824, USA
  \ead{vicente.leyton@physik.uni-hamburg.de}}

%
%

\begin{abstract}

  We investigate the quantum noise properties of a weakly nonlinear
  Duffing resonator in the deep quantum regime, where only few quanta
  are excited. This regime is dominated by the appearance of coherent
  multiphoton resonances in the nonlinear response of the resonator to
  the modulation. We determine simple expressions for the photon noise
  spectrum and find that the multiphoton resonances also induces a
  multiple peak structure in that noise.  When the corresponding
  multiphoton Rabi oscillations are underdamped, zero temperature
  quantum fluctuations determine comparable populations of all
  quasienergy states which belong to a resonant multiphoton
  doublet. Most interestingly, the quantum fluctuations probe the
  multiphoton transitions by inducing several peaks in the noise
  spectrum of the resonator observables. In particular, the noise of
  the photon number contains complete information about the
  multiphoton states and their stationary populations via pairs of
  nearly symmetric peaks at opposite frequencies.  Their widths are
  determined by the damping of the Rabi oscillations and their heights
  are proportional to the stationary nonequilbrium populations. A
  finite detuning from a multiphoton resonance generates a
  quasielastic noise peak at zero frequency. In addition, we relate
  the stationary populations of the quasienergy states with an
  effective quantum temperature and discuss the role of a finite
  temperature.
\end{abstract}
\maketitle

\section{Introduction}

Coupling a driven quantum mechanical oscillator to environmental
fluctuations allows the oscillator dynamics to reach a stationary
state. In the stationary state, energy is coherently absorbed from the
pump and leaks into the environment via random dissipative
transitions, which inevitably induce noise in the resonator. This
occurs even at zero temperature where only environmental zero-point
fluctuations (quantum noise) exist. The noise properties of a
nonlinear oscillator determine many fundamental nonequilbrium
phenomena such as quantum heating
\cite{Peano2010a,Peano2010b,Dykman2011} and quantum activation
\cite{Dykman1988}.

Nonlinear oscillators are used as basic elements for quantum state
detection or amplification. Examples of those are the Josephson
bifurcation amplifier
\cite{Siddiqi04,Siddiqi05,Lupascu07,Vijay09,Siddiqi10, Ong2011} and
the cavity bifurcation amplifier \cite{Vijay09,Metcalfe07}.  In this
context, the noise properties of the resonator, which is used as
detector or amplifier, determine the backaction of the measurement or
amplification on the system itself
\cite{Maklin01,Clerk10,Serban2010}. Clearly, it is desirable to keep
the backaction as weak as possible, while on the other hand, a
significant coupling of the amplification or measurement device to the
system is useful in order to achieve a sufficiently strong detection
or amplification efficiency. A fundamental lower limit for the
introduced disturbance, however, will be set by the quantum
noise. Hence, in order to design useful concepts for quantum state
detection based on nonlinear resonators in the deep quantum regime,
their quantum noise properties have to be addressed.

The Josephson bifurcation amplifier takes advantage of the dynamically
induced bistability due to the nonlinearity of the resonator.  The
eigenstates of the qubit are mapped onto the coexisting stable
vibrational states of the resonator, which have different amplitudes
and phase relations relative to the phase set by the external
drive. Hence, they allow for a large discrimination power. Up to
present, these amplifying devices mostly operate in a regime where
many quanta in the resonator are excited. This implies that pure
quantum fluctuations are typically small on average.  Nevertheless,
some experiments have been realized at low temperature where the
relevant fluctuations are quantum mechanical in nature
\cite{Vijay09,Ong2011}. The regime of weak fluctuations has been the
subject of intense theoretical investigation
\cite{Peano2010a,Peano2010b,Dykman2011,Serban2010, Drummond80,
  Drummond81,Dykman2012, Andre2012,Leyton11}. It has been shown, that
the stationary distribution over the quasienergy states of the driven
oscillator at zero temperature has the form of an effective Boltzman
distribution, thereby allowing to introduce the concept of an
effective quantum temperature implying quantum heating even at $T=0$.
Signatures of the onset of quantum fluctuations can be seen in the
relative intensities of the lines of the resonator noise spectrum
\cite{ Dykman2011,Drummond80,Drummond81, Dykman2012, Andre2012} and in
the appeareance of a fine structure in the spectral lines of
resonators with comparatively large nonlinearities and large quality
factors \cite{Dykman2011, Dykman2012}. The spectral fine structure
yields detailed information on the quasienergy distribution
\cite{Dykman2011, Dykman2012}.

Recently it has been proposed that nonlinear quantum detectors which
operate in the regime of few quanta (deep quantum regime) would bring
different advantages, such as a small back action, a large
discrimination power with an enhanced readout fidelity, and a
sufficiently large measurement efficiency \cite{Leyton11}.  In the
deep quantum regime, the frequency-resolved nonlinear response of the
oscillator to the external driving with frequency $\omega_{\rm ex}$
shows a rich fine structure
\cite{Peano2010a,Peano2010b,Peano04,Peano06a,Peano06b} which is mainly
generated by few-photon transitions in the resonator. The splitting of
the typical Lorentzian resonance of a harmonic oscillator into a
series of non-Lorentzian resonances and antiresonances reflects the
intrinsic nonequidistance of the energy levels $E_n$ of a nonlinear
oscillator.  $N$-photon transitions with the resonance condition
$E_N-E_0=N\hbar\omega_{\rm ex}$, $N=1,2,\dots$, and the subsequent
drift down along the ladder of the few-photon Fock states generate a
pronounced nonequilbrium quasienergy distribution which is strongly
different from the Boltzman-type
\cite{Peano2010a,Peano2010b,Peano06a}. Peaks or dips in the nonlinear
response are a direct consequence of the nonequilbrium distribution
over states with different oscillation amplitude and phase
\cite{Peano2010a,Peano2010b,Peano06a}.  The signatures of such a
characteristic non-Lorentzian lineshape of the response has been
observed experimentally in a circuit-cavity QED set-up \cite{Bishop}.

\begin{figure}[t]
  \centering
  \includegraphics*[width=\textwidth]{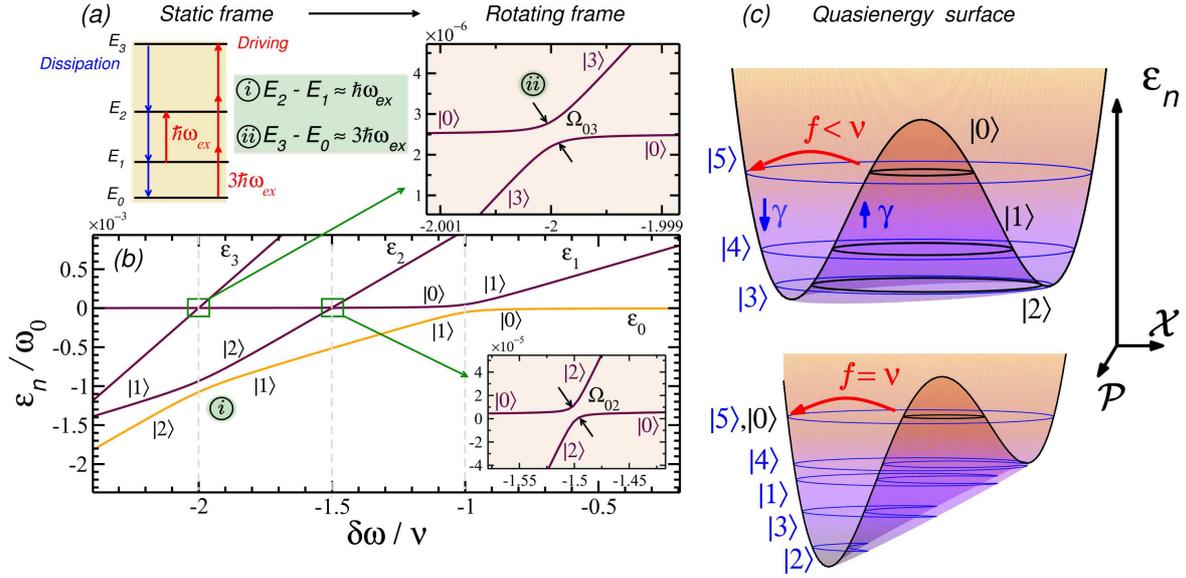} 
  \caption{Multiphoton Rabi transitions around the $N=3$-photon
    resonance $\delta\omega = \delta\omega_3$. In panel (a) we depict
    a sketch of the driving-induced resonant $3$-photon transitions
    (red arrows) in the nonlinear oscillator. Likewise, the blue
    arrows indicate the noise-induced relaxation process. In panel
    (b), we show the underlying quasienergy spectrum as a function of
    the external frequency together with two zooms to the avoided
    crossings for the $2-$ and $3-$photon Rabi transitions for $\nu =
    10^{-3}\omega_0$ and $f = \nu/10$. In panel (c), we schematically
    indicate of the coherent multiphoton Rabi transitions (red arrow)
    and the dissipative transitions (blue arrows) on the quasienergy
    surface which results from a semiclassical approach, see text. The
    upper figure shows the less tilted case when $f=\nu/10$, while for
    the lower figure $f=\nu$.  We emphasize that relaxational
    transitions at zero temperature typically occur in both
    directions, i.e., downwards {\em and\/} upwards along the
    quasienergy surface, which is in striking contrast to dissipative
    transitions in {\em static\/} potential surfaces, where only
    ``downward relaxation'' is possible.  An escape due to ``upward
    relaxation'' is known as quantum activation
    \cite{Dykman1988}.  \label{fig1}}
\end{figure}

In this work, we investigate the noise properties of modulated
nonlinear oscillators in the deep quantum regime. We consider the
simplest example of a monostable anharmonic oscillator which has a
quartic nonlinearity (Duffing oscillator).  Such a weakly nonlinear
Duffing oscillator has a remarkable symmetry: its energy levels $E_n$
with $n\leq N$ are pairwise resonant for the same driving frequency
$\omega_{\rm ex}$, $E_{N-n}-E_{n}=(N-2n)\hbar\omega_{\rm ex}$. An
example of the energy spectrum for the case $N=3$ is sketched in Fig.\
\ref{fig1} (a). After preparing the oscillator in its $n$-th excited
state $n\leq N$, it displays periodic quantum oscillations between the
$n$-th and the $N-n$-th excited states. During these oscillations,
$|N-2n|$ photons are being exchanged between the oscillator and the
modulation field. The oscillations of the photon number $\hat{n}$ are
usually referred to as multiphoton Rabi oscillations. Their
characteristic frequency, the Rabi frequency $\Omega_{nN}$, depends on
the intensity of the driving field and on the number of photons
exchanged. The Rabi frequency $\Omega_{0N}$ for the $N$-photon
oscillations is the smallest Rabi frequency. The multiphoton Rabi
oscillations with $N-n$ photons involved are underdamped if their Rabi
frequency $\Omega_{nN}$ exceeds the dissipative rate of photon leaking
into the environment. The latter is the oscillator relaxation rate
$\gamma$. For $\gamma\ll\Omega_{0N}$ all the Rabi oscillations are in
general underdamped.  The periodically driven resonator reaches its
stationary state on the timescale $\gamma^{-1}$.

In the stationary state, quantum noise induces -- even at zero
temperature -- fluctuations in the photon number $\hat{n}$. The
dynamics of these fluctuations is characterized by multiphoton
oscillations which manifest themselves as peaks in the noise spectrum
$S(\omega)$ of $\hat{n}$, located at plus/minus the Rabi frequencies
$\Omega_{nN}$. In the underdamped regime, the dissipative dynamics of
the driven oscillator is most appropriately described in terms of
random transitions between the oscillator quasienergy states. When the
driving is resonant, the pairs of oscillator Fock states with $n$- and
$N-n$-photons are resonantly superposed. The corresponding oscillator
quasienergy states are a symmetric and an antisymmetric superposition
of the two Fock states. Their splitting in quasienergy is given by the
Rabi frequency $\Omega_{nN}$.  The corresponding peak in the noise
spectrum at $(-)\Omega_{nN}$ is due to random transitions from the
state with (highest) lowest to that with the (lowest) highest
quasienergy of the doublet. The peak intensity is proportional to the
stationary occupation probability of the initial quasienergy state.
Therefore, the noise spectrum offers a convenient way to directly
probe the stationary distribution over all the quasienergy states.
Moreover, for weak driving and exactly zero detuning from the
multiphoton resonance, the noise spectrum of the $\hat{n}$-photon
transition is symmetric, i.e., $S(\omega)=S(-\omega)$ and two
inelastic peaks are signatures of an oscillatory decay of the
fluctuations towards the stationary state. States belonging to a
multiphoton doublet then have the same stationary occupation
probabilities.  For a weakly detuned modulation or a stronger driving,
the spectrum becomes asymmetric. Besides, an additional quasielastic
peak appears at zero frequency which represents incoherent relaxation
of the fluctuations towards the stationary state. These features have
some analogy in the spectral correlation function of a (static)
quantum mechanical two-level system weakly coupled to a dissipative
harmonic bath \cite{Weiss}. There, the spin correlation function is a
sum of three Lorentzians. The two inelastic peaks are symmetrically
located at finite frequencies and their width determines the inverse
of the dephasing time. In addition, the quasielastic peak at zero
frequency represents incoherent relaxation with the inverse relaxation
time given by its width. In the driven system, the appeareance of a
quasielastic peak depends on the intriguing interplay between the
nonlinearity, the driving strength and the dissipation strength.
 
\section{Multiphoton Rabi oscillations of the Duffing oscillator}

We consider a periodically modulated quantum oscillator with mass $ m
$, eigenfrequency $ \omega_0 $ and a quartic (Kerr) nonlinearity
described by the Hamiltonian
\begin{equation}\label{ham}
  H(t) = \frac{1}{2m}p^2 + \frac{1}{2}m \omega_0^2
  x^2 + \frac{1}{4}\alpha x^4 +F x 
  \cos(\omega_{\rm ex}t).
\end{equation}
The modulation amplitude $F$ is assumed to be so small that it induces
only weakly nonlinear vibrations. This is guaranteed by the condition
$\alpha A^2\ll m\omega_0^2$, with $A(F)$ being the typical amplitude
of the nonlinear vibrations.  The modulation frequency $\omega_{\rm
  ex}$ is chosen to be close to the oscillator eigenfrequency
$\omega_0$ such that the detuning $\delta \omega$ is small, i.e.,
\begin{equation}
 \delta\omega\ll \omega_0\,,\qquad \delta\omega\equiv \omega_0-\omega_{\rm ex}.
\end{equation} 
Our theory applies to hard as well as to soft nonlinearities
$\alpha\lessgtr 0$, but for concreteness we will focus on the case of
a hard nonlinearity, $\alpha>0$.

The quantum dynamics of the weakly detuned and weakly nonlinear driven
oscillator is most conveniently described in terms of the oscillator
ladder operators $a$ and $a^\dagger$, in a rotating frame determined
by the unitary transformation
\begin{equation}\label{eq:rotframe}
R(t)=\exp[-i\omega_{\rm ex}a^\dagger a t] \,.
\end{equation}
In the rotating frame, the typical time scale of the resonator
dynamics is given by $\delta\omega^{-1}$, so that terms oscillating
with frequencies $\pm2\omega_{\rm ex}$ and $\pm4\omega_{\rm ex}$
average out and can be neglected in the transformed Hamiltonian
$R(t)\,H(t)\,R^\dagger(t) - i \, \hbar R(t)\,\dot{R}^\dagger(t)$.
Thereby, we obtain the RWA Hamiltonian
\begin{equation} \label{eq:hamrwa} 
\tilde{ H} = \delta\omega \hat{n} +
  \nu \hat{n} (\hat{n} +1)/2 + f (a^\dagger + a)/2,
\end{equation}
where $\hat{n}\equiv a^\dagger a$ is the photon number operator, $\nu$
and $f$ are the frequencies associated with the Kerr nonlinearity and
the external field amplitude at the quantum scale $x_{\rm
  ZPF}=\sqrt{\hbar/m\omega_0}$, i.e., $\nu=3\alpha x_{\rm
  ZPF}^4/4\hbar$ and $f=Fx_{\rm ZPF}/\sqrt{2}\hbar$.  In order to keep
the notation compact we have set $\hbar=1$ in Eq.\ (\ref{eq:hamrwa})
and in the remainder of the paper.  The oscillator quasienergies
$\varepsilon_n$ and quasienergy states $|\psi_n\rangle$ are the
eigenvalues and eigenvectors of the rotating wave Hamiltonian,
$\tilde{H}|\psi_n\rangle=\varepsilon_n|\psi_n\rangle$.  For vanishing
driving, the quasienergy spectrum is given by
\begin{equation}\label{eq:spectrum}
\varepsilon_n=\delta\omega \, n + \nu n (n+1)/2\qquad {\rm for } \qquad  f\to 0
\, . 
\end{equation}
We are primarily interested in studying the noise spectrum in presence
of multiple multiphoton resonances $E_{N-n}-E_n=(N-2n)\omega_{\rm ex}$
for $ n\leq N$, or equivalently $\varepsilon_{N-n}=\varepsilon_n$ for
$f\to 0$.  From Eq.\ (\ref{eq:spectrum}) we find the resonant
condition
\begin{equation}\label{eq:Nresfreq}
 \delta\omega=-\nu (N+1)/2\equiv\delta\omega_N.
\end{equation}
Up to leading order in the driving, the quasienergy eigenstates
$|\psi_n\rangle$ for $n\leq N\neq N/2$ are a resonant superposition of
the pair of oscillator Fock states $|n\rangle$ and $|N-n\rangle$,
i.e.,
\begin{eqnarray}\label{eq:resstates}
 |\psi_n\rangle\approx\left(|n\rangle\pm|N-n\rangle\right)/\sqrt{2} \, .
\end{eqnarray}
We choose the signs $-$ and $+$ for $n<N/2$ and $N/2<n\leq N$,
respectively.  In the following, we refer to the resonant
superposition of a pair of Fock states as resonant states or
multiphoton states.  The states $|\psi_n\rangle$ which are not
involved in a multiphoton transition ($n>N$ and $n=N/2$ for $N$ even)
can be approximated as the corresponding Fock states
$|\psi_n\rangle\approx|n\rangle$.  The Rabi frequency $\Omega_{nN-n}$
of the multiphoton oscillations within the pair of Fock states
$|n\rangle$ and $|N-n\rangle$ is given by the splitting of the
corresponding levels $\varepsilon_n$ and $\varepsilon_{N-n}$
\cite{Larsen76,Peano06a} as
\begin{equation}\label{eq:Rabi}
 \Omega_{nN-n}=|\varepsilon_n-\varepsilon_{N-n}|= 
 f \left( \frac{f}{\nu} \right)^{N-2n-1}\frac{
{(N-n)!^{1/2}}}{{n!^{1/2}}(N-2n-1)!^2}
 \, .
\end{equation}
The resonance condition in Eq.\ (\ref{eq:Nresfreq}) is not
renormalized by a finite driving within the RWA. Only for a
comparatively larger driving $f\sim\nu\ll\omega_0$, the multiphoton
transitions have to be reinterpreted as tunneling transitions between
semiclassical states \cite{Dmitriev86,Dykman05}.

As we shall detail in Section \ref{powerspectrum}, the multiphoton
Rabi oscillations induce peaks in the spectral densities of oscillator
observables only when the Rabi frequency $\Omega_{0N}$ for the
multiphoton transition from the zero-photon ground state is larger
than the noise-indcued level broadening of the relevant quasienergy
levels $\varepsilon_0$ and $\varepsilon_N$. In the next section, we
will pave the wave for the calculation of the noise spectrum in this
regime, by formulating the master equation for a weakly nonlinear
oscillator and by evaluating the stationary occupation populations
over the quasienergy states.

\section{Stationary dynamics in   the deep quantum regime} 
\label{Sec:dissipative}

In the presence of a weak bilinear coupling to the fluctuations of a
bosonic bath, the assumptions of small detuning and weak nonlinearity
that underly the RWA naturally lead to a Liouville-von Neuman quantum
master equation in the Lindblad form for the density matrix
$\hat{\rho}$ of the weakly damped oscillator in the rotating frame
\cite{Peano06a,Peano06b},
\begin{eqnarray} 
\dot{\hat{\rho}} = {\cal L}\hat{\rho}&\equiv& 
-i[\tilde{H},{\hat{\rho}}]+\gamma(\bar{n}+1){\cal D}[a]\hat{\rho}+
\gamma \bar{n}{\cal D}[a^\dagger]\hat{\rho}\nonumber\\
{\cal D}[O]\hat{\rho}&\equiv& 
([O\hat{\rho},O^\dagger]+[O,\hat{\rho} O^\dagger])/2.\label{eq:Liouville}
\end{eqnarray}
Here, ${\cal L}$ and ${\cal D}[O]$ are the Liouville and the Lindblad
superoperators, respectively.  Moreover, $\gamma$ is the oscillator
damping rate for which we assume that $\gamma\ll\omega_0$. It results
from a standard Ohmic bath spectral density $J(\omega)=\gamma
\omega$. In addition, $\bar{n}$ is the bosonic occupation number at
frequency $\omega_0$ and temperature $T$ and is given by
$\bar{n}=(e^{\omega_0/k_BT}-1)^{-1}$.

\subsection{The stationary distribution }

For long times, the density matrix in the rotating frame $\hat{\rho}$
relaxes to a stationary state $\hat{\rho}^{\infty}$, satisfying
\begin{equation}\label{eq:stsol}
 {\cal L} \hat{\rho}^{\infty} = 0.
\end{equation}
When the oscillator decay rate $\gamma$ is larger than the driving,
$\gamma\gg f$, the width of the resonant quasienergy levels
$\varepsilon_n$ induced by the bath fluctuations are larger than the
corresponding Rabi frequency $\Omega_{nN-n}$ of the multiphoton
transitions.  Then, the multiphoton resonances are smeared out and the
coherent effects associated with multiphoton oscillations are strongly
suppressed.  Hence, dissipation sets a lower limit for the driving
strength, $f\gg\gamma$, which has to be overcome in order to observe
multiphoton transitions. On the other hand, for comparatively larger
driving $f\sim\nu$, quantum fluctuations are significantly reduced and
the oscillator is latched to a classical attractor at asymptotic times
which are much larger than the typical relaxational time scale,
$\gamma^{-1}$. In this regime, the (quasi)stationary distribution of
the population over the quasienergy states assumes Boltzman form
\cite{Peano2010a,Peano2010b,Dykman2011,Dykman2012,Andre2012}. \\
Here, we restrict our analysis to the deep quantum regime where the
driving is larger than the damping but smaller than the nonlinearity,
$\gamma\ll f \ll \nu$. Thereby, we have implicitly assumed a
comparatively large nonlinearity $\nu\gg\gamma$.

\subsubsection{Underdamped regime:}

We start our discussion assuming that all Rabi oscillations are
underdamped. Put differently, we assume that the smallest Rabi
frequency $\Omega_{0N}$ is larger than the relevant level broadening.
We refer to this regime as the fully underdamped regime. Then, the
off-diagonal matrix elements of $\hat{\rho}^\infty$ projected onto the
quasienergy basis $|\psi_l\rangle$ are negligible and we can set them
to zero, i.e., we perform a secular approximation,
\begin{equation}
  \rho^\infty_{lk}\equiv\langle\psi_l|\hat{\rho}^\infty|\psi_k\rangle=
  0\qquad {\rm for}\qquad l\neq k \, .
\end{equation}
Then, a balance equation for the stationary occupation probabilities
$\rho^\infty_{ll}$ follows from Eqs.\ (\ref{eq:Liouville}) and
(\ref{eq:stsol}) according to
\begin{equation}\label{eq:balanceeq}
 \gamma_l\rho^\infty_{ll}-\sum_{l\neq k}W_{l,k}\rho^\infty_{kk}=0 \, .
\end{equation}
Here, $W_{l,k}$ is the transition rate from state $|\psi_k\rangle$ to
state $|\psi_l\rangle$,
\begin{equation}
\qquad W_{l,k}\equiv\gamma\left[(1+\bar{n}) 
|\langle \psi_l|a|\psi_k\rangle|^2+\bar{n}|\langle
\psi_l|a^\dagger|\psi_k\rangle|^2\right]
\end{equation}
and $\gamma_l$ is the width of quasienergy level $\varepsilon_l$ given
by $\gamma_l\equiv\sum_{k\neq l}W_{l,k}$.  We can now formulate more
precisely the condition for underdamped Rabi oscillations to occur
within the pair forming the narrowest resonance, which is
$\Omega_{0N}\gg\gamma_0$.
 
The solution for stationary occupation probabilities up to leading
order in the small parameters $f/\nu$ and $\bar{n}$ is given in Ref.\
\cite{Peano06a}: The pair of multiphoton states $|\psi_{n}\rangle$ and
$|\psi_{N-n}\rangle$ in Eq.\ (\ref{eq:resstates}) have equal
stationary population, i.e.,
$\rho^\infty_{nn}=\rho^\infty_{N-n\,N-n}$. The pair with the narrowest
resonance has the occupation probabilities
$\rho^\infty_{00}=\rho^\infty_{NN}$. The occupation probability grows
algebraically with $n<N/2$ as
\begin{equation}\label{eq:relocc}
  \rho^\infty_{n+1\,n+1}= \frac{N-n}{n+1}\rho^\infty_{n\,n}\qquad  {\rm
for}\qquad n<N/2 \, .
\end{equation}
The states $|\psi_l\rangle$ with $l>N$ have vanishing occupation
probability, $\rho^\infty_{ll}= 0$.  As follows from the discussion
above, the degeneracy $\rho^\infty_{00}=\rho^\infty_{NN}$ is
approximate and is lifted for higher order in $f/\nu$.

\subsubsection{Quasienergy distribution close to a multiphoton
  resonance:}
\label{section:closetoresonance}

One can easily generalize the above expressions to the case where the
detuning $\delta\omega$ does not exactly match the resonant condition,
$\delta\omega\neq\delta\omega_N$. Since the Rabi frequencies for the
different pairs of resonant transitions in Eq.\ (\ref{eq:Rabi}) are
exponentially different, we can choose
$|\delta\omega-\delta\omega_N|\ll\Omega_{1N-1}$, so that all the pairs
of Fock states $|n\rangle$ and $|N-n\rangle$ with $1<n<N/2$ are still
resonantly superposed, except for
\begin{eqnarray} \label{dress1}
 |\psi_{0}\rangle = \cos \frac{\theta}{2} |0\rangle - \sin \frac{\theta}{2}
|N\rangle \ , \quad
{\rm and } \quad 
|\psi_{N}\rangle = \sin \frac{\theta}{2}|0\rangle + \cos \frac{\theta}{2} |N\rangle  \ , 
\end{eqnarray}
with $\theta =\tan^{-1}[\Omega_{0N}/N(\delta\omega-\delta\omega_N)]$.
The corresponding solution for the stationary density matrix close to
resonance is \cite{Peano06a}
\begin{eqnarray}
 &&\rho^\infty_{NN}=\rho^\infty_{00}\tan^4\frac{\theta}{2}\,,
 \qquad\rho^\infty_{11}=\rho^\infty_{00}N\tan^2\frac{\theta}{2}\, ,\nonumber\\ 
&&\rho^\infty_{n+1\,n+1}= \frac{N-n}{n+1}\rho^\infty_{n\,n}\qquad {\rm for}\qquad 1\leq n<N/2\,.
\label{eq:rhooutofresonance}
\end{eqnarray}

\subsubsection{Partially underdamped regime:}

Next we consider a comparatively large relaxation rate $\gamma$, so
that the narrowest Rabi resonance is overdamped but the remaining
resonances are still underdamped, $\Omega_{0N}\ll N\gamma\ll
\Omega_{1N-1}$.  We refer to this regime as the partially underdamped
regime.  Then, incoherent multiphoton transitions from the ground
state $|0\rangle$ to state $|N\rangle $ with a small rate
$\Omega_{0N}^2/(N \gamma)$ and the subsequent emission of excitations
into the bath determines a small but finite occupation of the resonant
states $\rho^\infty_{nn}$, $n\ge 1$.  Formally, the stationary
distribution $\hat{\rho}^\infty$ can be obtained by setting all the
off-diagonal elements of $\rho^\infty_{lk}$ to zero except for
$\rho^\infty_{N0}$ and $\rho^\infty_{0N}$ and solving Eq.\
(\ref{eq:stsol}). Thereby, we find
\begin{eqnarray}
 &&\langle0|\hat{\rho}^\infty|0\rangle\approx 1,\qquad 
\langle N|\hat{\rho}^\infty|N\rangle\approx \Omega_{0N}^2/(N^2\gamma^2)
\nonumber\\
 &&\rho^\infty_{11}=\Omega_{0N}^2/(N\gamma^2)\qquad {\rm for}
\qquad \Omega_{0N}^2/(N\gamma^2)\gg\exp[-\omega_0/(k_B T)]\nonumber\\ 
&&\rho^\infty_{n+1\,n+1}= \frac{N-n}{n+1}\rho^\infty_{n\,n}\qquad {\rm for}\qquad 1\leq n<N/2
\,.
\end{eqnarray}
The crossover between this solution and the fully underdamped solution
Eq.\ (\ref{eq:relocc}) is given in Ref.\ \cite{Peano06a}.  Both
stationary nonequilbrium distributions are determined by quantum
fluctuations and are very different from the equilibrium
Boltzmann-type distribution when a driven resonator is latched to a
classical attractor.

\subsection{The nonlinear response of the oscillator}

In the steady state regime, $t\gg\gamma^{-1}$, the oscillator state is
described by the time-independent density matrix $\hat{\rho}^\infty$
in the rotating frame and the oscillator dynamics is embedded in the
time-dependent reference frame $R(t)$. The mean value of an
observables $O$ is
\begin{equation}
 \langle O(t)\rangle_\infty\equiv\lim_{t\to\infty}\langle O(t)\rangle=  
 \mathrm{Tr}[\hat{\rho}^\infty R^\dagger(t)OR(t)]\,.
\end{equation}
Therefore, the stationary oscillations of the position expectation
value $\langle x(t)\rangle_\infty$ are sinusoidal,
\begin{equation}\label{eq:nlinres}
 \langle x(t)\rangle_{\rm \infty}=
\sqrt{2}x_{\rm ZPF}\cos{(\omega_{\rm ex}t+\varphi)}|\langle a \rangle_\infty|,\qquad
\langle a \rangle_\infty=\sum_{lk}\rho^\infty_{lk}\langle\psi_l|a|\psi_k\rangle
\, .
\end{equation}
It has been shown that the nonlinear response $\langle
x(t)\rangle_{\rm \infty}$ of the oscillator as a function of
$\omega_{\rm ex}$ shows resonances and antiresonances in the deep
quantum regime \cite{Peano04,Peano06a,Peano06b}.  The response is
proportional to the transmitted amplitude in a heterodyne measurement
scheme and it has already been measured for a weakly nonlinear
oscillator \cite{Bishop}.  Clearly, such a measurement scheme, or more
general, any measurment scheme which probes stationary mean values as
opposed to correlations does not allow to resolve the different
degenerate resonances separately. Neither, they allow us to access the
stationary distribution $\rho^\infty_{ll}$ directly. This becomes
possible only when correlations, e.g., via noise spectra are
measured. In the next section, we show that this can indeed be
achieved by measuring the spectrum of the photon number noise.

\section{The noise  spectrum in the deep quantum regime}
\label{powerspectrum}

\subsection{Definition of the noise spectrum}

The Lindblad master equation (\ref{eq:Liouville}) in general also
allows to investigate transient phenomena and correlation functions.
Its formal solution for a given initial state $\hat{\rho}_0$ is given
by $\hat{\rho}(t)=e^{{\cal L} t}\hat{\rho}_0$.  Moreover, a general
correlator $\langle O'(t') O(t) \rangle$ can be evaluated as the mean
value of the operator $O'$ at time $t'$ with the virtual operator
$R^\dagger(t)OR(t) \rho(t)$ at time $t$. This view has been
established several decades ago by the Lax formula
\cite{Melvin60,Melvin63} according to
\begin{equation}
  \langle O'(t') O(t) \rangle = 
  \mathrm{Tr} \lbrace R^\dagger(t')O'R(t') e^{{\cal L}(t'-t)}R^\dagger(t)OR(t) \rho(t) \rbrace
\end{equation}
For long times $t\gg \gamma^{-1}$, we find that    
\begin{eqnarray}\label{eq:correlator}
&&\langle O(t+\delta t) O'(t) \rangle_{\rm \infty}\equiv 
\lim_{t\to\infty} \langle O(t+\delta t) O'(t) \rangle\nonumber\\
&&=\mathrm{Tr} 
\lbrace R^\dagger(t+\delta t)OR(t+\delta t) e^{{\cal L} \delta t} R^\dagger(t)O'R(t) 
\rho^{\rm \infty} \rbrace\,.
\end{eqnarray}
In general, such correlators are periodic functions of the preparation
time $t$. The noise spectrum is defined as a double average over
quantum fluctuations and the time $t$.  \\

Here, we are specifically interested in the noise spectrum $S(\omega)$
of the autocorrelator of the photon number $\hat{n}$, $\langle
\hat{n}( t+\delta t) \hat{n}(t) \rangle_{\rm \infty}$.  From Eqs.\
(\ref{eq:rotframe}) and (\ref{eq:correlator}), we find
\begin{eqnarray}\label{eq:nncorrelator}
&&\langle \hat{n}(t+\delta t) \hat{n}(t) \rangle_{\rm \infty}=\mathrm{Tr} 
\lbrace \hat{n} e^{{\cal L} \delta t} \hat{n} \rho^{\rm \infty} \rbrace\,.
\end{eqnarray}
Since this correlator does not depend on the initial time $t$ as a
consequence of the RWA, we can define the noise spectrum in terms of a
single average over quantum fluctuations according to
\begin{equation}\label{eq:emissionspectrum}
  S(\omega) = 2 \, \mathrm{Re} \int_0^{\infty} d  t \, e^{i \omega t}\langle \hat{n}( t)
\hat{n}(0) \rangle_{\rm \infty}. 
\end{equation}
It is useful to separate the contributions to $S(\omega)$ into those
coming from the expectation value of $\hat{n}$, and those from its
fluctuations, i.e.,
\begin{eqnarray}
  S(\omega) &= &\langle \hat{n}\rangle_{\rm \infty}^2\delta(\omega)+\delta S(\omega),\nonumber\\
\delta S(\omega)&\equiv& 2 \, \mathrm{Re} \int_0^{\infty} d  t \, e^{i \omega t}\langle \delta\hat{n}( t)
\delta\hat{n}(0) \rangle_{\rm \infty}. \label{eq:emissionspectrum0}
\end{eqnarray}
Here, $\delta \hat{n}$ is the operator for the photon number
fluctuations, i.e., $\delta n=\hat{n}-\langle \hat{n}\rangle_{\rm
  \infty}$.
 
Our path to compute the noise spectrum consists in three steps: i) We
express the virtual preparation $\hat{n}\hat{\rho}^\infty$ in terms of
right eigenvectors of the superoperator ${\cal L}$. ii) We plug the
resulting decomposition into Eq.\ (\ref{eq:nncorrelator}). Then, each
term decays exponentially with a different exponent which is given by
the corresponding eigenalue of ${\cal L}$. iii) We compute the Fourier
integral in Eq.\ (\ref{eq:emissionspectrum}), which thereby yields a
sum over (overlapping) Lorentzians.

The general expression, which is useful for a concrete numerical
evaluation, for the noise spectrum given in terms of the eigenvectors
and the eigenvalues of ${\cal L}$ is derived in \ref{AppendixA}. In
the next section, we consider the special case of underdamped
multiphoton Rabi oscillations.

\subsection{Noise spectrum in the  underdamped regime}

When all the multiphoton Rabi oscillations are underdamped,
$\Omega_{N0}\gg \Gamma_N$, the coherences
$|\psi_{N-n}\rangle\langle\psi_{n}|$ and
$|\psi_{n}\rangle\langle\psi_{N-n}|$ are approximate eigenvectors of
the Liouvillian ${\cal L}$. Then,
\begin{eqnarray}
 {\cal L}|\psi_n\rangle\langle\psi_{N-n}|&=&
-(\Gamma_n-i\Omega_{nN-n})|\psi_n\rangle\langle\psi_{N-n}|\quad {\rm for} \quad
n<N/2\,,
\nonumber\\
{\cal L}|\psi_{N-n}\rangle\langle\psi_{n}|&=&
-(\Gamma_n+i\Omega_{nN-n})|\psi_{N-n}\rangle\langle\psi_{n}|\quad {\rm for}
\quad n<N/2\, .
\label{eq:righteig}
\end{eqnarray}
with the level widths being given as
$\Gamma_n=\gamma_n=\gamma(\bar{n}+1/2)N+\gamma\bar{n}$ for
$n<(N-1)/2$.  For $N$ odd,
$\Gamma_{(N-1)/2}=\gamma(1+2\bar{n})(5N+1)/8+\gamma\bar{n}$.  Up to
leading order in $f/\nu$, the decomposition of the virtual preparation
$\hat{n}\hat{\rho}^\infty$ in terms of right eigenvectors of ${\cal
  L}$ has the simple expression
\begin{equation}\label{eq:inprep}
 \hat{n} \hat{\rho}^\infty\approx \left(N/2\right) \hat{\rho}^\infty -\sum_{n<N/2}(N/2-n)\rho^\infty_{nn}
\left(|\psi_n\rangle\langle\psi_{N-n}|+|\psi_{N-n}\rangle\langle\psi_{n}|\right)\,.
\end{equation}
Clearly, each term of the above decomposition yields a Lorentzian peak
in the noise spectrum $S(\omega)$.  The first term yields the
contribution to $S(\omega)$ from the expectation value of $\hat{n}$,
$(N/2)^2\delta(\omega)$.  The remaining terms yield inelastic peaks
associated to random transitions between quasienergy states belonging
to the same multiphoton doublet. Since the populations
$\rho^\infty_{nn}$ and $\rho^\infty_{N-nN-n}$ are approximately equal,
peaks at opposite frequency have approximately equal intensity. By
putting together Eqs.\ (\ref{eq:nncorrelator}),
(\ref{eq:emissionspectrum}), (\ref{eq:righteig}), and
(\ref{eq:inprep}), we find $S(\omega)=(N/2)^2\delta(\omega)+\delta
S(\omega)$ with
\begin{eqnarray}\label{eq:powerspectrmulti}
 \delta S(\omega)\approx\sum_{n<N/2} S_n(\omega)+S_{N-n}(\omega),&&\\
 S_n(\omega)=S_{N-n}(-\omega)=
\frac{2\Gamma_n\rho_{nn}^{\infty}(N/2-n)^2}{(\omega-\Omega_{nN-n})^2+
\Gamma_n^2}\,.&&
\end{eqnarray} 
Hence, the Lorentzians are centered at the multiphoton Rabi
frequencies $\Omega_{nN-n}$ and have a resonance width of
$\Gamma_n$. The factor $(N-2n)^2/4$ is the leading order expression
for the squared matrix element
$|\langle\psi_n|\hat{n}|\psi_{N-n}\rangle|^2$.  Remarkably, the line
intensities depend only weakly on the driving $f$ and on the
temperature through the stationary distribution $\rho^\infty_{nn}$.
Up to leading order, the driving $f$ enters only in the splitting of
the lines through the Rabi frequencies.  Notice that Eq.\
(\ref{eq:powerspectrmulti}) is valid only in the vicinity of a
multiphoton peak since terms of order $\gamma$ are not taken into
account. In order to evaluate the tails of the peaks more precisely,
one has to take into account the contribution stemming from all
eigenvectors of ${\cal L}$, see \ref{AppendixA}.

In the left and right panels of Fig.\ \ref{fig2}, we show the noise
spectrum $S(\omega)$ for the cases $N=2$ and $N=3$, respectively.  The
noise spectrum for $N=2$ shows a pair of symmetric peaks which
correspond to the transitions
$|\psi_0\rangle\leftrightarrow|\psi_{2}\rangle$. Likewise, the noise
spectrum for $N=3$ displays two pairs of symmetric peaks corresponding
to the transitions $|\psi_0\rangle\leftrightarrow|\psi_{3}\rangle$ and
$|\psi_1\rangle\leftrightarrow|\psi_{2}\rangle$. The green dashed
lines mark the results from our approximate analytical formula in Eq.\
(\ref{eq:powerspectrmulti}) while the yellow solid lines show the data
obtained by numerically evaluating the expression in Eq.\
(\ref{eq:powerspectr1}).  An excellent agreement is found.

In Fig.\ \ref{fig2}a), additional smaller side peaks of the order of
$f/\nu$ are also visible, see the gray lines representing a ten-fold
zoom. They are not associated to any resonant transition between
multiphoton states and are thus not captured by the leading order
expression given in Eq.\ (\ref{eq:powerspectrmulti}). The particular
subleading peaks in Fig.\ \ref{fig2}a) belong to the transitions
$|\psi_0\rangle\leftrightarrow|1\rangle$.

\begin{figure}[t]
  \centering
  \includegraphics*[width=\textwidth]{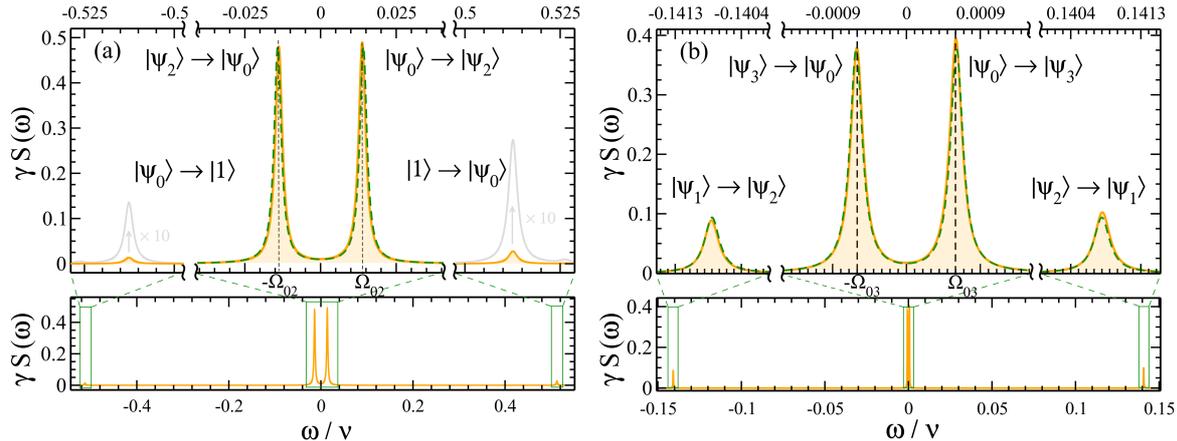} 
  \caption{Photon noise spectrum at the $N=2$- and the $N=3$-photon
    resonance are shown in panel (a) and (b) respectively for $\nu =
    10^{-3}\omega_0$, $f=\nu/10$, $\gamma = \Omega_{02}/10$ in (a) and
    $\gamma =\Omega_{03}/10$ in (b). Shown are the approximate results
    obtained with Eq. (\ref{eq:powerspectrmulti}) (dashed green
    lines), and the results from a full numerical solution of the
    general expression for the spectrum derived in Appendix A (orange
    solid lines). The gray lines in panel (a) mark a zoom to the
    subleading off-resonant transitions. \label{fig2}}
\end{figure}  

These features have a direct analogy in the spectral correlation
function of a static quantum mechanical two-level system which is
weakly coupled to a dissipative harmonic bath \cite{Weiss}. For a
general biased two-state system with anticrossing energy levels, the
pair correlation function is a sum of three Lorentzians. The two
inelastic peaks are symmetrically located at finite frequencies and
their width determines the inverse of the dephasing time. For a biased
static two-level system away from resonance, an additional
quasielastic peak at zero frequency appears which represents
incoherent relaxation with the inverse relaxation time given by its
width. Since we consider here the case strictly at resonance (in the
RWA), no zero-frequency peak is present.

\subsection{Photon anti-bunching}

In general, the photon emission characteristics of a quantum
mechanical resonator can show peculiar nonclassical features. For
instance, counterintuitive correlation phenomena such has photon
antibunching can occur, where the photon number correlation function
for short delay times is smaller than the one for classical,
uncorrelated photons.  This implies that the probability for photons
to arrive in pairs is suppressed \cite{Walls2008}. Our approach
provides a natural framework to investigate a possible non-Poissonian
statistics of the multiphoton events in the nonlinear
resonator. Therefore, we consider the normalized photon number
correlation function or second-order coherence function defined as
\begin{equation}
  g^{(2)}(\tau) = \frac{\langle  a^\dagger(t)a^\dagger(t+\tau)a(t+\tau)
    a(t)\rangle_\infty}{\langle a^\dagger(t)a(t) \rangle_\infty \langle
    a^\dagger(t+\tau)
    a(t+\tau) \rangle_\infty} \, . 
\end{equation}
For long delay times $\tau$, the counts of two photons with a delay
time $\tau$ are statistically independent events, $g^{(2)}(\tau \to
\infty)=1$. For vanishing delay times, we have
\begin{equation}
  g^{(2)} (\tau \to 0) = 1+
  \frac{\langle \hat{n}^2\rangle_\infty-\langle \hat{n}\rangle_\infty^2-
    \langle \hat{n}\rangle_\infty}{\langle n\rangle_\infty^2}\, . 
\end{equation}
Photon antibunching corresponds to the case $ g^{(2)} (\tau =
0)<1$. For the fully underdamped case, we find the expression
\begin{equation}
 g^{(2)} (\tau = 0) = 
  \frac{2N (N-1)+4\sum_{1}^{N-1}n(n-N)\rho_{nn}}{N^2}=1-\frac{1}{N}\ , 
\end{equation}
which represents the known result of the second-order correlation
function of the electromagnetic field \cite{Walls2008}.  Hence, the
oscillator displays photon antibunching close to a multiphoton
transition. The second-order coherence of the stationary state of the
quantum Duffing oscillator at the $N$-th multiphoton resonance has the
same value as the second order coherence for an oscillator prepared in
the single Fock state $|N\rangle$, in spite of its fluctuations over
the quasienergy states.

\subsection{Lineshape of the noise spectrum close to a multiphoton
  resonance}

In presence of a small detuning from the multiphoton resonance,
$\delta\omega-\delta\omega_N\sim\Omega_{0N}$, the states
$|\psi_0\rangle$ and $|\psi_N\rangle$ are no longer a resonant
superposition of the Fock states $|0\rangle$ and $|N\rangle$.  Hence,
the corresponding stationary occupation probabilities $\rho_{00}$ and
$\rho_{NN}$, given in Eq.\ (\ref{eq:rhooutofresonance}), become
significantly different. In turn, the pair of peaks $S_0(\omega)$ and
$S_N(\omega)$, which are associated to the transitions
$|\psi_0\rangle\leftrightarrow|\psi_{N}\rangle$, become asymmetric
such that $S_0(\omega)\neq S_N(-\omega)$.  This behavior is shown in
shown in Fig.\ \ref{fig4}a) for the case around the $3$-photon
resonance.  The peak lineshapes can readily been evaluated and we find
\begin{equation}
 S_{0}(\omega)=
\frac{2\Gamma_0\rho_{00}^{\infty}N^2(\sin\theta\cos\theta)^2}{(\omega-\varepsilon_N+\varepsilon_0)^2
+\Gamma_0^2},\qquad 
S_{N}(\omega)=
\frac{2\Gamma_N\rho_{NN}^{\infty}N^2(\sin\theta\cos\theta)^2}{(\omega-\varepsilon_0+\varepsilon_N)^2
+\Gamma_N^2}\,.
\end{equation}
Their distance increases with the quasienergy splitting,
$\varepsilon_N-\varepsilon_0=\mathrm{sgn}(\delta\omega-\delta\omega_N)
(\Omega_{0N}^2+N^2|\delta\omega-\delta\omega_N|^2)^{1/2}$, whereas the
peak width does not change close to the multiphoton resonance,
$\delta\omega-\delta\omega_N\sim\Omega_{0N}$. The asymmetry is
determined by the stationary occupation probabilities
$\rho^\infty_{00}$ and $\rho^\infty_{NN}$.  From Eq.\
(\ref{eq:rhooutofresonance}), we find
\begin{equation}\label{eq:asymm} 
 \frac{S(\omega)}{S(-\omega)}=\frac{\rho^\infty_{00}}{\rho^\infty_{NN}}=
\cot^4\frac{\theta}{2}=
  \left[\frac{\Omega_{0N}}{|\varepsilon_N-\varepsilon_0|-N(\delta\omega -
\delta\omega_N)} \right]^4 \,.
\end{equation}
The above expression is valid for $\omega$ close to the center of the
largest peak, $\omega\sim\varepsilon_N-\varepsilon_0$, and
$|\delta\omega-\delta\omega_N|$ not too large such that
$S(\pm\omega)\gg\gamma$.

In addition to the peaks at finite frequencies (which induce decaying
coherent multiphoton Rabi oscillations), also a zero frequency peak
appears. This quasielastic peak is associated to incoherent
relaxational decay of the multiphoton Rabi oscillations and is also
known for the noise correlation function of a static biased quantum
two level system \cite{Weiss}.

\begin{figure}[t]
  \centering
  \includegraphics*[width=\textwidth]{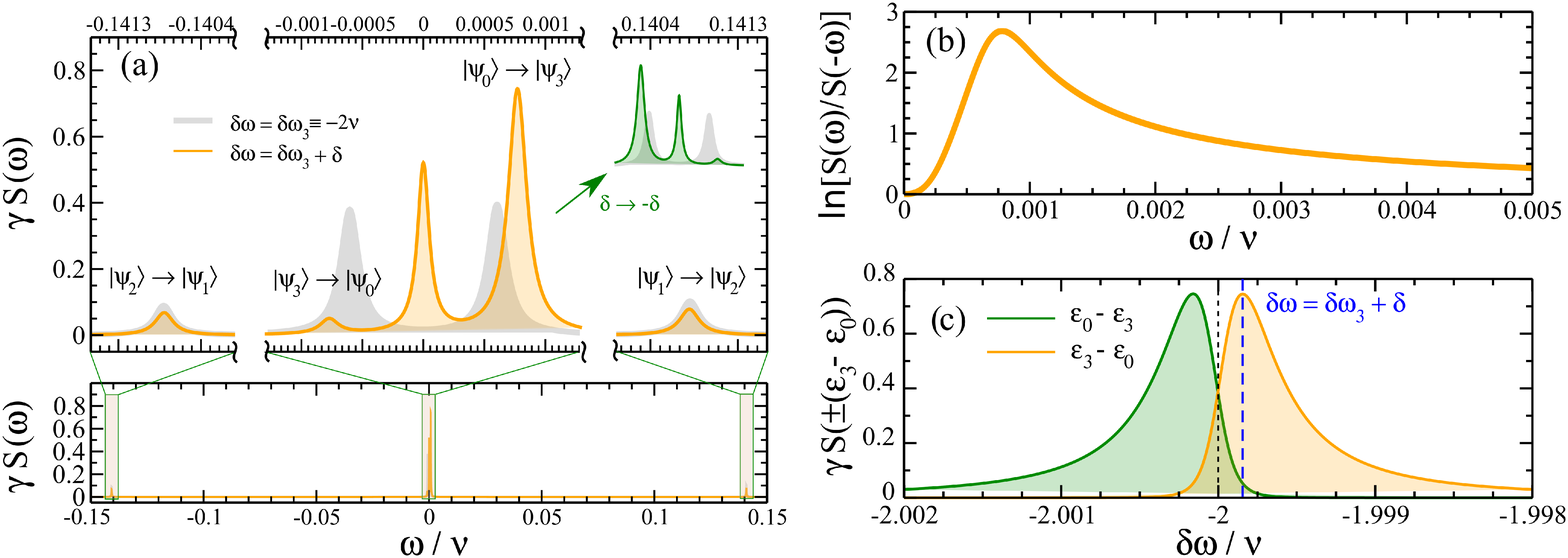}
  \caption{(a) Asymmetric structure of the photon noise spectrum at
    frequency $\delta\omega = \delta\omega_3 + \delta $, i.e., out of
    resonance for a detuning $\delta = 1.6\times10^{-4}\nu$ for the
    same parameters used in Fig.\ \ref{fig2}b) (orange solid line). In
    addition, we show in the background the symmetric photon noise at
    the resonant frequency $\delta\omega_3$ (grey shadowed
    area). Moreover, we depict the inverted case $\delta \rightarrow
    -\delta$, which shows a symmetric behavior under the reflection
    $\omega \rightarrow -\omega$ (green solid line).  (b) Noise
    asymmetry via the logarithm of Eq.\ (\ref{eq:asymm}) for the same
    parameters as in (a). (c) Height of the photon noise peak for the
    transition $|\psi_0\rangle \rightarrow |\psi_3\rangle$ (orange
    solid line), and $|\psi_3\rangle \rightarrow |\psi_0\rangle$
    (green solid line) as a function of the external frequency. The
    peak maximum is located at $\delta \omega_3 \pm
    \delta$. \label{fig3}}
\end{figure}

In Fig.\ \ref{fig3}b), we show the logarithm of the asymmetry ratio
given in Eq.\ (\ref{eq:asymm}).  The asymmetry shows a clear maximum
at approximately $\varepsilon_3-\varepsilon_0$.

To further illustrate the asymmetry in the peak heights, we show in
Fig.\ \ref{fig3}c) the peak maxima associated to the transitions
$|\psi_0\rangle \rightarrow |\psi_3\rangle$ and $|\psi_3\rangle
\rightarrow |\psi_0\rangle$. At the $3$-photon resonance (black dashed
vertical line), both peaks are equal in height (symmetric noise
spectrum). Away from the resonance, the low (high) frequency branch
aquires more spectral weight for negative (positive) detuning.

\subsection{Photon noise at zero frequency}
\label{Sec:zerofreqnoise}

\begin{figure}[t]
  \centering
  \includegraphics*[width=85mm]{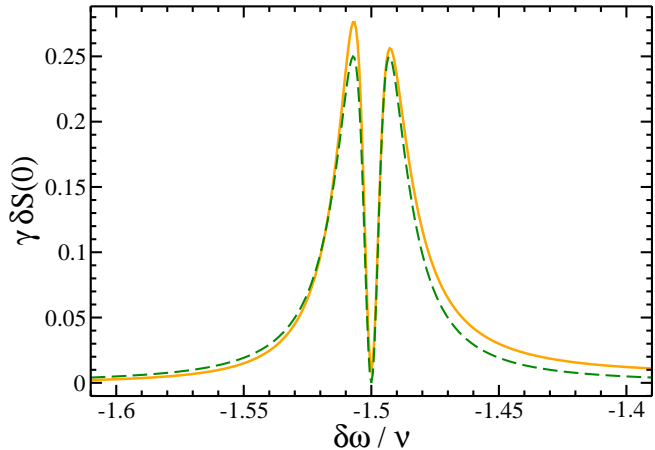}
  \caption{Photon noise at zero probe frequency as a function of the
    external frequency for the parameter set used in
    Fig. \ref{fig2}(a) evaluated around the second multiphoton
    resonance, $\delta\omega \sim -3\nu/2$.  Shown is the comparison
    of the approach Eq. (\ref{PS0}) as a green dashed line with the
    numerical simulation as a orange solid line. \label{fig4}}
\end{figure}

Fluctuations of an oscillator (quasi)energy induce a broad (with width
$\propto\gamma$) zero frequency peak in the noise spectrum of an
observable whose mean value depends on the (quasi)energy
\cite{Dykman1988b}.  For weak driving $f\ll\nu$ and at a resonance
$|\delta\omega-\delta\omega_N|\ll\Omega_{0N}$, the quasienergy states
of the Duffing oscillator have large fluctuations as several
quasienergy states have comparable occupation probabilities even at
$T=0$. However, the mean value of $\hat{n}$ becomes independent from
the quasienergy, $\langle\psi_n|\hat{n}|\psi_n\rangle\approx N/2$ for
$n\leq N$.  As a consequence, the contribution to the noise spectrum
of $\hat{n}$ coming from fluctuations $\delta S(\omega)$ does not have
a peak at zero frequency since $\delta S(0)\propto \gamma$. Close to
resonance, when $|\delta\omega-\delta\omega_N|\sim\Omega_{0N}$, two
dynamical effects compete: on one hand, the quasienergy fluctuations
quickly decrease for increasing detuning, i.e., moving away from
resonance as the occupation probability of the state $|\psi_0\rangle$
approaches one.  On the other hand, the mean value of $\hat{n}$
becomes strongly dependent on the quasienergy.  As a result of this
competition, the intensity of the zero frequency noise plotted as a
function of $\delta\omega$ has two maxima at the two opposite sides of
the resonant value $\delta\omega_N$. In Fig.\ \ref{fig4}, we show the
zero frequency noise for the special case $N=2$.  The yellow solid
line represents the intensity at zero frequency computed numerically,
while the green dashed line is the leading order contribution (in
$f/\nu$)
\begin{equation}\label{PS0}
  \delta S(\omega = 0) \approx 
  4\gamma^{-1}\sin^2 \frac{\theta}{2}\cos^2 \frac{\theta}{2}
  \left(\sin^2 \frac{\theta}{2}-\cos^2
    \frac{\theta}{2}\right)^2 \ . 
\end{equation}

\subsection{Noise spectrum towards the semiclassical regime}

\begin{figure}[t]
  \centering
  \includegraphics*[width=\textwidth]{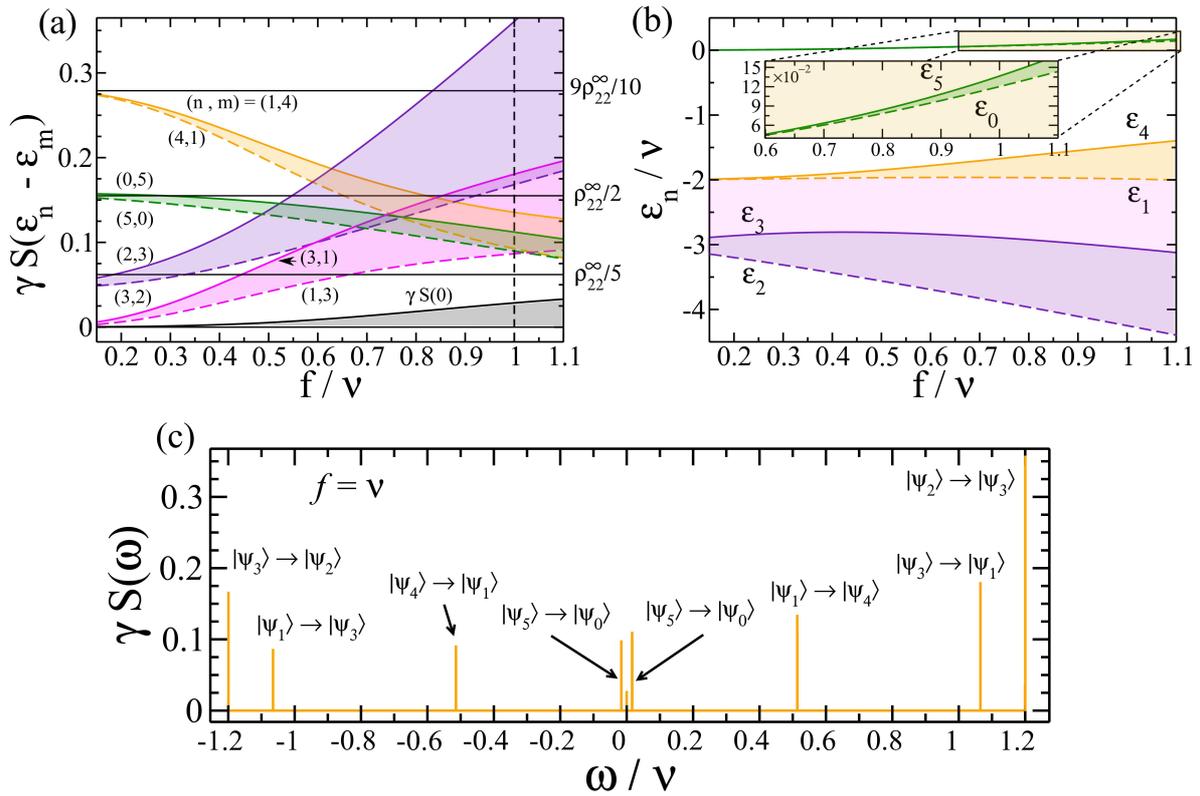}
  \caption{(a) Height of the photon noise peaks corresponding to the
    transitions withiin the pair $|\psi_n\rangle \leftrightarrow
    |\psi_{m}\rangle$ as a function of the driving strength $f$. Each
    pair is marked by a solid and a dashed line in the same color. In
    addition, we depict the rise of the zero frequency peak (black
    solid line) as the driving strength increases. The black
    horizontal lines indicate the expected values of the noise level
    evaluated up to leading order in $f/\nu$ by using Eq.\
    (\ref{eq:powerspectrmulti}). The parameters are $\nu =
    10^{-3}\omega_0$, $\delta\omega = \delta\omega_5$.  In panel (b),
    we show for the same parameters the quasienergy spectrum as a
    function of the driving strength $f$. In panel (c), the photon
    noise spectrum as a function of the probe frequency $\omega$ is
    shown for a large driving strength $f=\nu$.\label{fig5}}
\end{figure}

Next, we investigate the noise spectrum for larger driving strengths,
$f\lesssim \nu$.  In order to illustrate how the noise spectrum
changes for increasing driving, we show the intensities of the
brightest peaks as a function of the driving strength for the
$N=5$-photon resonance in ; see Fig.\ \ref{fig5}a).  In Fig.\
\ref{fig5}b), we also show the quasienergy spectrum, and the noise
spectrum for a comparatively large value of the driving amplitude
$f=\nu$ is shown in Fig.\ \ref{fig5}c). A peak in the noise spectrum
at frequency $\omega=\varepsilon_l-\varepsilon_k$ is associated to a
single transition $|\psi_k\rangle\rightarrow|\psi_l\rangle$ and is
given by
\begin{equation}
  S(\omega)=\sum_{lk}\frac{2\rho_{ll}^{\infty}|\langle \psi_l |a^\dagger
    a|\psi_k\rangle|^2(\gamma(a_l-a_k)^2+\Gamma_l+\Gamma_k)}{
    (\omega+\varepsilon_l-\varepsilon_k)^2+(\gamma(a_l-a_k)^2+\Gamma_l+\Gamma_k)^2}
  \, .
\end{equation}
Hence, the relative intensities of a pair of peaks at opposite
frequencies is still related to the occupation probability of the
corresponding initial states through
$S(\varepsilon_l-\varepsilon_k)/S(\varepsilon_k-\varepsilon_l)=\rho_{kk}/\rho_{
  ll } $.

For weak driving, we have three pairs of approximately symmetric peaks
as described by Eq.\ (\ref{eq:powerspectrmulti}). Each peak
corresponds to a transition between two states belonging to a
multiphoton doublet of quasidegenerate states:
$|\psi_0\rangle\leftrightarrow|\psi_5\rangle$,
$|\psi_1\rangle\leftrightarrow|\psi_4\rangle$, and
$|\psi_2\rangle\leftrightarrow|\psi_3\rangle$. For increasing driving,
the spectrum becomes increasingly asymmetric.  For moderate values of
the driving, the noise spectrum undergoes two major qualitative
changes: i) the peak at zero frequency becomes clearly visible; ii) a
pair of peaks corresponding to the transitions
$|\psi_1\rangle\leftrightarrow|\psi_3\rangle$ acquires a significant
intensity. For $f=\nu$, the peak associated with the transition
$|\psi_3\rangle \rightarrow|\psi_1\rangle$ is even the second
brightest peak.

These qualitative changes can be explained in terms of a semiclassical
description valid beyond the weak driving limit.  The RWA Hamiltonian
in Eq.\ (\ref{eq:hamrwa}) can be rewritten in terms of rotating
quadratures, and interpreted as a quasienergy surface in phase space
\cite{Dmitriev86,Dykman05}.  It has the shape of a tilted mexican hat
and is sketched in Fig.\ \ref{fig1} (c) for two values of $f$. The
larger $f$ is, the stronger is the induced tilt. The local maximum and
the minimum of the quasienergy surface are the classical
attractors. In the static frame, they describe stationary oscillations
with a small and a large amplitude, respectively. In the vicinity of
the attractors the vibration amplitude and the slow part of the
oscillation phase display slow vibrations with frequency
$\propto\delta\omega$. In absence of resonant transitions, each
quasienergy state can be associated to a quantized quasiclassical
orbit which lies on the internal surface around the local maximum, on
the external surface, or along the quasienergy well around the
minimum. For very weak driving, $f\ll \nu/\sqrt{2(N+1)}$, the quantum
mechanical Fock states $|n\rangle$ with $n<N/2$ are associated to
quasiclassical trajectories on the internal surface around the local
maximum, whereas the Fock states with photon number $n$ larger than
$N/2$ are associated to semiclassical orbits on the external
surface. Within this representation, the multiphoton transitions can
then be reinterpreted as tunneling transitions between the internal
and the external parts of the surface \cite{Dmitriev86,Dykman05}. For
comparatively larger driving, the zero-point quasienergy associated to
the slow vibrations around the minimum ($\propto\delta\omega$) becomes
smaller than the dynamical barrier height. Then, quasienergy states
appear which are localized in the quasienergy well. In turn, the noise
spectrum becomes qualitatively different from the one for weak
driving. The small quantum fluctuations around the minimum of the
quasienergy surface can be described in terms of an effective
auxiliary oscillator with ladder operators $b$ and $b^\dagger$ and are
given by
 \begin{equation}
 a=a_h+b\cosh{r^*_h}-b^\dagger\sinh{r^*_h}\,. 
\label{oldcreation}
\end{equation}
Here, $a_h$ is the amplitude of the stationary oscillations rescaled
by $\sqrt{2}x_{\rm ZPF}$ \cite{Dykman2012,Andre2012}. They can be
mimicked by a local effective quantum temperature $T_e=(2 k_B \ln
\coth r^*_h)^{-1}$ which depends on the squeezing factor $r^*_h$
\cite{Peano2010a,Peano2010b,Dykman2012,Andre2012}.  For $f=\nu$, the
states $|\psi_2\rangle$, $|\psi_3\rangle$, and $|\psi_1\rangle$ can be
identified with the groundstate and first two excited states of the
auxiliary oscillator (but in the remainder of this discussion we keep
the same labels for the states as in the weak driving limit). The
level spacing $\varepsilon_3-\varepsilon_2$ is of the order of the
frequency of the slow classical oscillations of the amplitude and slow
part of the phase.

Such oscillations appear in the noise spectral density of a classical
oscillator as a pair of peaks.  In a nonlinear quantum oscillator
whose quasienergy levels are not equidistant and their distance
exceeds the damping strength, the classical peaks have a ``quantum''
fine structure \cite{Dykman2011}.  In the present case of the Duffing
oscillator, the classical noise peak is splitted into two peaks
associated to the nearest neighbor transitions between the ground
state and the first excited state, and the first and the second
excited state, $|\psi_2\rangle\leftrightarrow|\psi_3\rangle$ and
$|\psi_3\rangle\leftrightarrow|\psi_1\rangle$, respectively. Their
peak height is proportional to the square of the rescaled vibration
amplitude $a_h$ and to the occupation of the initial state
$\rho^\infty_{nn}$. The latter, in particular, is governed by the
quantum temperature $T_e$. For the ratio of the peak heights, we find
\cite{Dykman2011}
\begin{equation}
\frac{ S(\varepsilon_3-\varepsilon_2)}{S(\varepsilon_2-\varepsilon_2)}\approx
\frac{\rho^\infty_{22}}{\rho^\infty_{33}}\approx\coth^2r^*_h\approx\frac{\rho^\infty_{33}}{\rho^\infty_{11}}
\approx \frac{
S(\varepsilon_1-\varepsilon_3)}{S(\varepsilon_3-\varepsilon_1)}.
\end{equation}
Next nearest neighbor transitions can also yield peaks in the noise
spectra of a Duffing oscillator \cite{Andre2012}. In the present case,
the transitions $|\psi_2\rangle\leftrightarrow|\psi_1\rangle$ yield a
pair of dimmer peaks, however, located at frequencies outside the
frequency range shown in Fig.\ \ref{fig5}.

In the weak damping, weak driving regime discussed so far, the
quasienergy well around the minimum is still very shallow, and the
oscillator can escape from the small amplitude attractor via
tunneling. Therefore, the oscillator is not latched to any of the
attractors and the noise spectral density has also peaks which are
associated to intrawell transitions. In particular, the pair of peaks
with the smallest splitting describes coherent tunneling oscillations
between the internal and the external part of the quasienergy surface
(coherent dynamical tunneling or multiphoton Rabi oscillations).

Before closing this section, we mention that for the stronger driving
$f=\nu$, also a zero frequency peak appears in the noise spectrum, see
Fig.  \ref{fig5}c), although the frequency detuning has been fixed to
the $5$-photon resonance $\delta \omega = \delta\omega_5$. However, as
discussed above, this resonance condition is only valid for small
$f\ll \nu$, which is obviously not fullfilled. So the larger driving
induces an effective small detuning away from the exact avoided
quasienergy level crossing and generates an effective bias. Then, a
relaxation pole appears in the relevant self energy \cite{Weiss} which
corresponds to a quasielastic relaxation peak at zero frequency.

\begin{figure}[t]
  \centering
  \includegraphics*[width=\textwidth]{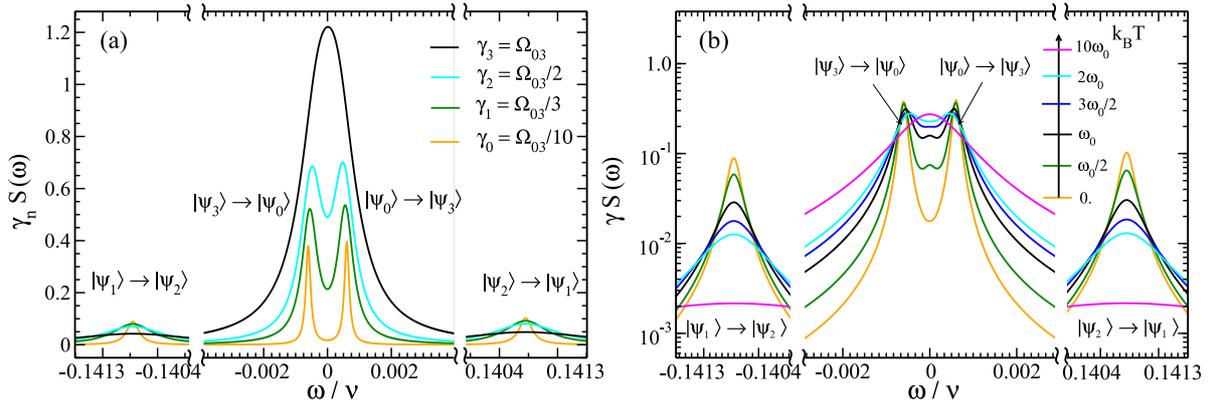}
  \caption{Photon noise at the third multiphoton resonance,
    $\delta\omega = \delta\omega_3$, as a function of the probe
    frequency. In panel (a), we show the behavior when going from the coherent
to the
    incoherent regime by increasing the damping constant from
    $\gamma_0 \ll \Omega_{03}$ to $\gamma_3 = \Omega_{03}$ (at
    $T=0$). In panel (b), we fix the damping constant to
    $\gamma = \Omega_{03}/10$ and  show the noise temperatures increasing 
from zero ($T=0$) up to 
    finite temperatures ($T \gg\omega_0$). The remaining parameters are 
$\nu=10^{-3}\omega_0$ and $f=\nu/10$. \label{fig6}}
\end{figure}  

\subsection{Dependence of the noise spectrum on damping and
  temperature}

So far, we have analyzed the case of zero temperature and small
damping, $\bar{n}\ll1$ and $\gamma\ll\Omega_{0N}$.  In this section,
we briefly address how the noise spectrum is modified for larger
damping and finite temperature by presenting numerical results of the
spectrum in a broad parameter range.

In Fig.\ \ref{fig6}a), we show $S(\omega)$ for different values of the
damping for the $3-$photon resonance where
$\delta\omega=\delta\omega_3$. As expected, the peaks in the noise
spectrum get broader for increasing damping.  Outside the fully
underdamped regime, the two peaks of the pair associated with the
transitions $|\psi_0\rangle\leftrightarrow|\psi_{3}\rangle$ start to
overlap and eventually merge into a single peak at zero frequency.
Thereby, the zero frequency noise is no longer suppressed
$S(\omega\approx 0)\propto\gamma^{-1}$, since incoherent relaxation
prevails over coherent decay for large damping. The peaks associated
with the underdamped transitions
$|\psi_1\rangle\leftrightarrow|\psi_{2}\rangle$ are still described by
Eq.\ (\ref{eq:powerspectrmulti}), even when the spectrum has a peak at
zero frequency. The decrease in the peak intensities reflects the
decrease of the populations $\rho^\infty_{11}$ and $\rho^\infty_{22}$
in the partially underdamped regime.

The dependence of the noise spectrum on temperature is shown in Fig.\
\ref{fig6}b) and behaves qualitatively similarly. For small
temperatures $\bar{n}\ll 1$, the spectrum is described by Eq.\
(\ref{eq:powerspectrmulti}). The temperature dependence enters in the
line widths of the quasienergy levels as well as in the stationary
distribution $\rho^\infty_{nn}$.  For larger temperatures, the two
low-frequency peaks merge into a single peak at zero frequency and the
side peaks becomes increasing broader as expected.

\section{Conclusions} 

In recent years, the rich phenomenology of driven and damped nonlinear
quantum oscillators has been impressively consolidated, including
their nonlinear response behavior in form of resonant and antiresonant
amplification, quantum coherent multiphoton Rabi oscillations, quantum
activation and quantum heating.  Gradually, the nontrivial effects
visible in noise correlation functions have also moved to the focus of
interest. Those become relevant whenever a nonlinear quantum
oscillator is used as a central element in an amplifier or quantum
measurement device.  In this work, we have analyzed the noise
properties of the quantum Duffing oscillator in the regine when only
few quanta are excited. Then, the nonlinear response shows pronounced
multiphoton peaks which are associated to resonant multiphoton Rabi
oscillations. The noise properties of these multiphoton transitions
show a rich phenomenology. To obtain the noise spectrum by analytical
means, we invoke the Lax formula for the autocorrelation function of
the photon number at different times and calculate its Fourier
transform. Exactly at a multiphoton resonance, the noise spectrum
consists in a collection of pairs of related resonances which are
located at opposite frequencies and which are equal in height.  Each
pair is associated to a multiphoton doublet. In spite of large
fluctuations over the oscillator quasienergy, no quasielastic peak
occurs at zero frequency. This is a consequence of a special symmetry
of the quantum Duffing oscillator: all quasienergy states which are
associated to a multiphoton doublet have the same mean value of the
photon number $\hat{n}$.

Slightly away from a multiphoton resonance, the noise spectrum becomes
asymmetric and the two resonances are no longer equal in height.  In
addition, as the mean values of $\hat{n}$ become different for
quasienergy states with comparable occupations, the quasielastic peak
emerges. Since the quasienergy fluctuations are suppressed away from a
multiphoton resonance, the intensity of the quasielastic peak as a
function of the detuning displays a maximum at the two opposite sides
of the resonant value $\delta\omega_N$.

Our approach also allows us to evaluate the transition to the
semiclassical regime by increasing the photon number by a larger
driving amplitude. Then, a quasiclassical quasipotential landscape in
phase space is a convenient tool to understand the stationary
nonequilibrium dynamics. This view directly leads to quantum
mechanical squeezed states which exist close to the local minimum of
the quasienergy landscape. A harmonic expansion allows us to
characterize the quantum fluctuations via an effective quantum
temperature. At larger (real) temperature and damping strengths, all
these quantum coherent features are washed out.

Although the time-resolved detection of noise properties of quantum
observables of driven resonators requires considerably more
experimental effort, we are confident that future experiments will
soon elucidate the importance of quantum noise in these systems.

\section*{Acknowledgements}
This work was supported by the DAAD (German Academic Exchange Service)
Research Grant No. Ref: A/08/73659 and by the DFG SFB 925 {\em Light
  induced dynamics and control of correlated quantum systems} (Project
C8).  V.P. was supported by the NSF (Grant No. EMT/QIS 082985). We
thank P. Nalbach, M.\ Marthaler and M.\ Dykman for valuable
discussions.
\appendix
\section{Numerical evaluation of the noise spectrum}
\label{AppendixA}
In order to numerically compute the noise spectrum $S(\omega)$, we
diagonalize the Liouvillian superoperator ${\cal L}$,
\begin{equation}
 {\cal L}\hat{k}_R=\lambda_k\hat{k}_R,\qquad   \hat{k}_L {\cal L}=\lambda_k\hat{k}_L
\end{equation}
Here, $\{\hat{k}_L\}$ and $\{\hat{k}_R\}$ are the sets of the left and
right eigenvectors of ${\cal L}$, respectively.  They constitute a
biorthogonal system implying that
$\mathrm{tr}\lbrace\hat{k}^\dagger_L\hat{k}'_R\rbrace=\delta_{kk'}$. The
corresponding eigenvalues $\lambda_k$ are in general complex.  The
stationary density matrix $\hat{\rho}^\infty$ is the only eigenvector
with zero eigenvalue.  All other eigenvectors represents transient
dissipative processes. The corresponding eigenvalues have a negative
real part. Note that we have implicitly assumed that ${\cal L}$ admits
a spectral decomposition. Put differently, we have ruled out that any
Jordan block has dimension larger than one. \\
With this at hand, the noise spectrum in Eqs.\
(\ref{eq:emissionspectrum}) can be rewritten in terms of $\hat{k}_L$,
$\hat{k}_R$ and $\lambda_k$ as a sum of partial spectra characterized
by Lorentzians according to
\begin{equation}\label{eq:powerspectr1}
  S(\omega)=\frac{2\sum_k(\mathrm{tr}\lbrace \hat{n}\hat{k}_R\rbrace 
    \mathrm{tr}\lbrace \hat{k}_L^\dagger\hat{n}\hat{\rho}^\infty\rbrace){\rm Re}\lambda_k}
  {(\omega -{\rm Im} \lambda_k)^2+({\rm Re} \lambda_k)^2}\,.
\end{equation}
The sum extends over those eigenvectors $\hat{k}_R$ which belong to
non-zero eigenvalues $\lambda_k$. Thereby, we have not included the
elastic Rayleigh peak $\langle
\hat{n}\rangle_{\infty}^2\delta(\omega)$ which trivially comes from
the stationary state $\hat{k}_R=\hat{\rho}^\infty$.

\end{document}